\documentclass[preprint2]{emulateapj-rtx4}
\usepackage{graphicx,times,bm}

\newcommand{\EQ}{\begin{equation}}
\newcommand{\EN}{\end{equation}}
\newcommand{\EQA}{\begin{eqnarray}}
\newcommand{\ENA}{\end{eqnarray}}

\newcommand{\Eq}[1]{Equation~(\ref{#1})}

\newcommand{\Fig}[1]{Figure~\ref{#1}}

\newcommand{\Tab}[1]{Table~\ref{#1}}

\newcommand{\bra}[1]{\langle #1\rangle}

{}
{}
{}

{}
{}
{}
{}
{}
{}
{}
{}
{}
{}
{}
{}
{}
{}
{}
{}
{}
{}
{}
{}

{}
{}
{}

{}

{}
{}

\newcommand{\kk}{\bm{k}}

\newcommand{\UU}{\mbox{\boldmath $U$} {}}

\newcommand{\BB}{\mbox{\boldmath $B$} {}}

\newcommand{\JJ}{\mbox{\boldmath $J$} {}}

\newcommand{\AAA}{\mbox{\boldmath $A$} {}}

\newcommand{\ff}{\mbox{\boldmath $f$} {}}

\newcommand{\WW}{\mbox{\boldmath $W$} {}}

\newcommand{\nab}{\mbox{\boldmath $\nabla$} {}}

\newcommand{\SSSS}{\mbox{\boldmath ${\sf S}$} {}}

\newcommand{\dd}{{\rm d} {}}

\newcommand{\const}{{\rm const}  {}}

\def\ga{\mathrel{\mathchoice {\vcenter{\offinterlineskip\halign{\hfil
$\displaystyle##$\hfil\cr>\cr\sim\cr}}}
{\vcenter{\offinterlineskip\halign{\hfil$\textstyle##$\hfil\cr>\cr\sim\cr}}}
{\vcenter{\offinterlineskip\halign{\hfil$\scriptstyle##$\hfil\cr>\cr\sim\cr}}}
{\vcenter{\offinterlineskip\halign{\hfil$\scriptscriptstyle##$\hfil\cr>\cr\sim\cr}}}}}
\def\Pm{\mbox{\rm Pr}_M}
\def\Rm{\mbox{\rm Re}_M}

\def\Rmc{R_{\rm m,{\rm crit}}}
\def\Rey{\mbox{\rm Re}}

\def\cs{c_{\rm s}}

\def\kf{k_{\it f}}

\def\epsK{\epsilon_{\it K}}
\def\epsM{\epsilon_{\it M}}
\def\epsT{\epsilon_{\it T}}
\def\EK{E_{\it K}}
\def\EM{E_{\it M}}
\def\kK{k_{\it K}}
\def\kM{k_{\it M}}

\def\Brms{B_{\rm rms}}

\def\urms{u_{\rm rms}}

\def\Beq{B_{\rm eq}}

\def\half{{\textstyle{1\over2}}}

\newcommand{\yapj}[3]{ #1, {ApJ,} {#2}, #3}

\newcommand{\yapjl}[3]{ #1, {ApJ,} {#2}, #3}

\newcommand{\yan}[3]{ #1, {Astron.\ Nachr.,} {#2}, #3}

\newcommand{\yana}[3]{ #1, {A\&A,} {#2}, #3}

\newcommand{\yass}[3]{ #1, {Ap\&SS,} {#2}, #3}
\newcommand{\ygafd}[3]{ #1, {Geophys.\ Astrophys.\ Fluid Dyn.,} {#2}, #3}

\newcommand{\yjfm}[3]{ #1, {J.\ Fluid Mech.,} {#2}, #3}

\newcommand{\ypf}[3]{ #1, {Phys.\ Fluids,} {#2}, #3}

\newcommand{\ypp}[3]{ #1, {Phys.\ Plasmas,} {#2}, #3}

\newcommand{\yprl}[3]{ #1, {Phys.\ Rev.\ Lett.,} {#2}, #3}

\newcommand{\ypre}[3]{ #1, {Phys.\ Rev.\ E,} {#2}, #3}

\newcommand{\yspd}[3]{ #1, {Sov.\ Phys.\ Dokl.,} {#2}, #3}

\newcommand{\yjour}[4]{ #1, {#2}, {#3}, #4}

\begin{document}
\title{Nonlinear small-scale dynamos at low magnetic Prandtl numbers}
\author{Axel Brandenburg}

\affil{
NORDITA, AlbaNova University Center,
Roslagstullsbacken 23, SE-10691 Stockholm, Sweden; and\\
Department of Astronomy, Stockholm University, SE-10691 Stockholm, Sweden
}

\email{brandenb@nordita.org
 ($ $Revision: 1.43 $ $)
}

\begin{abstract}
Saturated small-scale dynamo solutions driven by isotropic non-helical
turbulence are presented at low magnetic Prandtl numbers $\Pm$ down to 0.01.
For $\Pm<0.1$, most of the energy is dissipated via Joule heat and,
in agreement with earlier results for helical large-scale dynamos,
kinetic energy dissipation is shown to diminish proportional to
$\Pm^{1/2}$ down to values of 0.1.
In agreement with earlier work, there is, in addition to a short Golitsyn
$k^{-11/3}$ spectrum near the resistive scale also some evidence for a
short $k^{-1}$ spectrum on larger scales.
The rms magnetic field strength of the small-scale dynamo
is found to depend only weakly on the value of $\Pm$ and
decreases by about a factor of 2 as $\Pm$ is decreased from 1 to 0.01.
The possibility of dynamo action at $\Pm=0.1$ in the nonlinear regime is
argued to be a consequence of a suppression of the bottleneck seen in the
kinetic energy spectrum in the absence of a dynamo and, more generally,
a suppression of kinetic energy near the dissipation wavenumber.
\end{abstract}


\section{Introduction}

In astrophysical turbulence, dissipation of kinetic and magnetic
energies tends to occur on length scales much shorter than the scale of the
energy-carrying eddies.
Even though both kinetic and magnetic dissipation scales are comparatively
short, the current indications are that
it does matter which of the two is the shorter one and by how much.
Their ratio is the magnetic Prandtl number, $\Pm$.
For stars and liquid metals we have $\Pm\ll1$, while for galaxies $\Pm\gg1$.
An important example where the value of $\Pm$ is believed to matter
is the small-scale dynamo that converts kinetic turbulent energy into
magnetic energy under isotropic conditions.

A dynamo is only possible when the energy conversion is efficient
and larger than the magnetic energy dissipation.
This is quantified by the magnetic Reynolds number, $\Rm$, which is
a nondimensional measure of the inverse magnetic dissipation rate.
The critical value of $\Rm$, above which dynamo action occurs,
is known to increase with decreasing values of $\Pm$
\citep{RK97,BC04,HBD04,Scheko04,Scheko05,Scheko07,Iska07}.
In the following we define the magnetic Reynolds number as
$\Rm=\urms/\eta\kf$, where $\urms$ is the rms velocity fluctuation
of the turbulence, $\eta$ is the magnetic diffusivity,
and $\kf$ is the wavenumber of the energy-carrying eddies,
i.e., the wavenumber where energy is injected into the system.
The critical value of $\Rm$ is then found to be
around 35 for $\Pm=1$ and around 100 for $\Pm=0.2$, but note that
for \cite{Scheko04,Scheko05,Scheko07} and \cite{Iska07} the values of $\Rm$
are defined such that they are about 1.5 times larger than those
used here or in \cite{HBD04}.
For $\Pm=0.1$, however, no small-scale dynamo action has yet been found.
This may easily be a limitation of not having been able to increase
the fluid Reynolds number, $\Rey=\Rm/\Pm$, beyond 2000,
which limits $\Rm$ to 200 for $\Pm=0.1$ \citep{Iska07,Scheko07}.
Larger values of $\Rey$ have been possible by using hyperviscosity,
giving access to larger values of $\Rey$ and smaller values of $\Pm$
for fixed $\Rm$.
In that case, \cite{Iska07} and \cite{Scheko07} found small-scale
dynamo action for $\Pm=0.05$ and $\Rm=150$, i.e., the dynamo is now
easier to excite than for $\Pm=0.1$.
The reason for this is believed to be connected with the fact that the
properties of small-scale dynamos depend on the kinetic energy spectrum at
the resistive scale.
For $\Pm=1$, this scale is the viscous scale where the velocity
field is smooth in the sense that the velocity difference $\delta u$
over a separation $\delta\ell$ scales linearly, i.e., $\delta u\sim\delta\ell$.
For $\Pm\ll1$, following the argument of \cite{BC04}, the resistive
scale falls in the inertial range where the velocity field is rough and
$\delta u\sim\delta\ell^\zeta$ with $\zeta\approx0.4$, so the velocity
field would not be differentiable, making dynamo action inefficient.
However, for $\Pm=0.1$, the kinetic energy spectrum is even shallower
than in the inertial range, so the local value of $\zeta$ is even smaller
and the velocity field rougher than in the inertial range.
This phenomenon is known as the bottleneck effect \citep{Fal94,Kan03,Dob03}.
This bottleneck effect is believed to be the reason why $\Rmc$ reaches a
maximum at $\Pm\approx0.1$.

The usage of hyperviscosity does exaggerate the bottleneck, which still
exists even for the regular viscosity operator.
It would therefore be useful to verify small-scale dynamo action for
small values of $\Pm$ using the regular viscosity operator.
This will be done in the present paper.
In addition, we shall consider here the nonlinear regime, which has the
advantage that at small values of $\Pm$, much of the kinetic energy is
diverted to magnetic energy before it is dissipated viscously.
This allows one to increase the fluid Reynolds number beyond the maximal
value that would normally be possible at a given resolution.
This has been demonstrated in the context of helicity-driven
large-scale dynamos \citep{B09}, whose onset conditions are essentially
independent of the value of $\Pm$ \citep{B01,B09,Min07}.
This is not the case for the small-scale dynamos considered here,
where the flow is statistically isotropic and non-helical.

\begin{figure*}[t!]\begin{center}
\centering\includegraphics[width=\textwidth]{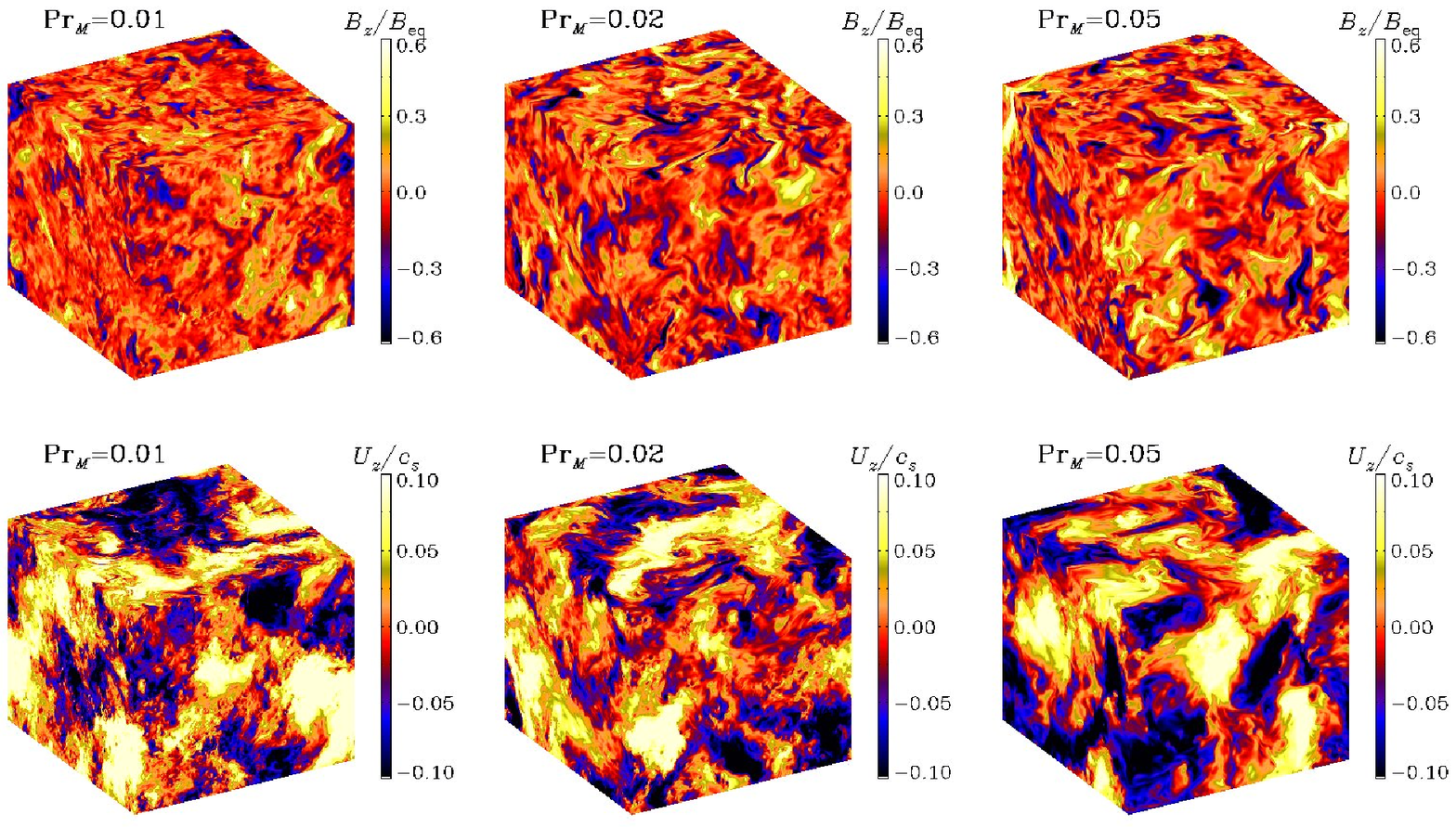}
\end{center}\caption[]{
Visualizations of $B_z$ and $U_z$ for $\Pm=0.01$ (left), $\Pm=0.02$,
and $\Pm=0.05$.
All runs are for $\Rm\approx160$ using $512^3$ mesh points.
}\label{UB}\end{figure*}

Our strategy for reaching low values of $\Pm$ is the same as that of
\cite{B09}.
We start with a simulation of a saturated small-scale dynamo at $\Pm=1$
and then increase the value of $\Rey$ while keeping the value of $\Rm$
in the range 150--160, provided the dynamo is still excited.
We are here particularly interested in the dependence of the saturation
field strength on $\Pm$ and the dissipation rate.

\section{The model}

Our model is similar to that presented in \cite{B01,B09} and
\cite{HBD03,HBD04}, where we solve the hydromagnetic equations for velocity
$\UU$, density $\rho$, and magnetic vector potential $\AAA$,
in the presence of an externally imposed
non-helical forcing function $\ff$,
for an isothermal gas with constant sound speed $\cs$, i.e.,
\EQ
{\partial\UU\over\partial t}=-\UU\cdot\nab\UU-\cs^2\nab\ln\rho
+\ff+(\JJ\times\BB+\nab\cdot2\rho\nu\SSSS)/\!\rho,\,
\label{dUU}
\EN
\EQ
{\partial\ln\rho\over\partial t}=-\UU\cdot\nab\ln\rho-\nab\cdot\UU,
\label{dlnrho}
\EN
\EQ
{\partial\AAA\over\partial t}=\UU\times\BB-\eta\mu_0\JJ.
\label{indEq}
\EN
Here, 
${\sf S}_{ij}={1\over2}(U_{i,j}+U_{j,i})-{1\over3}\delta_{ij}\nab\cdot\UU$
is the traceless rate of strain tensor, $\nu$ is the kinematic viscosity,
$\BB=\nab\times\AAA$ is the magnetic field, $\JJ=\nab\times\BB/\mu_0$
is the current density, and $\mu_0$ is the vacuum permeability.
We consider a triply-periodic domain of size $L^3$, so the smallest
wavenumber in the domain is $k_1=2\pi/L$.
The forcing function consists of plane waves with wavevectors $\kk$
whose lengths lie in the range $1\leq|\kk|/k_1\leq2$ with an average of
$\kf\approx1.5\,k_1$.
The amplitude of $\ff$ is such that the Mach number is $\urms/\cs\approx0.1$,
so compressive effects are negligible \citep{Dob03}.

Unless a simulation has been restarted from a previous one at another
value of $\Pm$, we start with a weak Gaussian distributed field in all
three components of $\AAA$, zero initial velocity, and uniform initial
density, $\rho=\rho_0=\const$, so the volume-averaged density remains
constant, i.e., $\bra{\rho}=\rho_0$.

In our simulations we vary the fluid Reynolds number and
the magnetic Prandtl number,
\EQ
\Rey=\urms/\nu\kf,\quad\Pm=\nu/\eta,
\EN
such that $\Rm=\urms/\eta\kf$ is in the range 150--160.
We also present a few results for $\Rm$ around 220.
We monitor the resulting kinetic and
magnetic energy dissipation rates per unit volume,
\EQ
\epsK=\bra{2\nu\rho\SSSS^2},\quad
\epsM=\bra{\eta\mu_0\JJ^2},
\label{epsdef}
\EN
whose sum, $\epsT=\epsK+\epsM$, is the total dissipation rate.
We use the fully compressible {\sc Pencil Code}\footnote{
http://www.pencil-code.googlecode.com} for all our calculations.
We recall that, for the periodic boundary conditions under consideration,
$\bra{2\SSSS^2}=\bra{\WW^2}+{4\over3}\bra{(\nab\cdot\UU)^2}$,
highlighting thus the analogy between vorticity $\WW=\nab\times\UU$ and $\JJ$
in the incompressible and weakly compressible cases.

\section{Results}

\subsection{Small-scale magnetic and velocity features at low $\Pm$}

In \Fig{UB} we present visualizations of $B_z$ and $U_z$ on
the periphery of the domain for three runs with $\Pm$ ranging
form 0.01 to 0.05.
Even though the value of $\Rm$ is the same in all three runs,
the magnetic field seems to have smaller scale structures in
the low $\Pm$ case.
The appearance of smaller scale structures is particularly clear
in the visualization of the velocity field for $\Pm=0.01$.

\begin{table*}[t!]\caption{
Summary of runs for different values of $\Pm$ and $\Rm\approx150$--$160$.
}\vspace{12pt}\centerline{\begin{tabular}{lcccccccccccc}
$\Pm$ & $\Rm$ & $\Brms/\Beq$ & $C_\epsilon$ & $\epsK/\epsT$ & $\epsM/\epsT$ &
$\kK$ & $\kM$ & $\Delta t/\tau$ & Res.\\
\hline
0.01&163&$0.12\pm0.02$&$0.34\pm0.03$&$0.49\pm0.13$&$0.68\pm0.06$&656&28&397&$512^3$\\
0.02&163&$0.15\pm0.04$&$0.44\pm0.03$&$0.60\pm0.24$&$0.64\pm0.10$&425&29&394&$512^3$\\
0.05&157&$0.28\pm0.03$&$0.65\pm0.04$&$0.31\pm0.08$&$0.77\pm0.05$&217&31&201&$512^3$\\
0.10&158&$0.28\pm0.04$&$0.65\pm0.01$&$0.39\pm0.07$&$0.72\pm0.04$&132&31&261&$256^3$\\
0.20&152&$0.32\pm0.04$&$0.73\pm0.10$&$0.50\pm0.13$&$0.67\pm0.05$& 85&30&147&$256^3$\\
0.50&150&$0.39\pm0.05$&$0.88\pm0.03$&$0.59\pm0.09$&$0.63\pm0.04$& 45&30& 98&$256^3$\\
1.00&146&$0.39\pm0.03$&$0.92\pm0.03$&$0.85\pm0.11$&$0.54\pm0.03$& 28&29&228&$256^3$
\label{Tsum}\end{tabular}}\end{table*}

In \Tab{Tsum} we summarize some essential properties of the simulations
for a sequence of simulations with different values of $\Pm$ between
0.01 and 1, but similar values of $\Rm$ of around 150--160.
The rms field strength relative to the equipartition value,
$\Beq=\urms\sqrt{\rho_0\mu_0}$, is about 0.3 for $0.05\leq\Pm\leq0.2$, while
for the runs with $\Pm=0.02$ and 0.01 it is about 0.15 and 0.12, respectively.
This is still a remarkably weak dependence that was not expected based on the
earlier results by \cite{Iska07} and \cite{Scheko07} for the onset conditions
of the small-scale dynamo.
As $\Pm$ is decreased from 1 to 0.01, $\epsK$ decreases and $\epsM$ increases.
However, the runs for $\Rm=160$ are rather close to the onset of dynamo action.
This becomes clear when comparing with two other runs for $\Rm=220$
and $\Pm=0.1$ and 0.02; see \Tab{Tsum2}.
For $\Pm=0.1$, $\Brms/\Beq\approx0.32$ and the ratio $\epsK/\epsT$
has dropped from 0.39 to 0.24, while for $\Pm=0.02$, $\Brms/\Beq\approx0.34$
and the ratio $\epsK/\epsT$ has dropped from 0.6 to 0.08.
Thus, we see that for values of $\Rm$ that are not too close to the onset
of dynamo action, the $\Pm$ dependence of $\Brms/\Beq$ is negligible and
$\epsK$ continues to drop.

The magnetic dissipation wavenumber, $\kM=(\epsM/\eta^3)^{1/4}$, is about 30
for all runs with $\Rm\approx150$--$160$, while the kinetic dissipation wavenumber,
$\kK=(\epsK/\nu^3)^{1/4}$, increases gradually with decreasing values
of $\Pm$ (or increasing values of $\Rey$).

\begin{table}[b!]\caption{
Comparison of runs for $\Rm\approx220$ and two values of $\Pm$.
}\vspace{12pt}\centerline{\begin{tabular}{lcccccccccccc}
$\Pm$ & $\Brms/\Beq$ & $C_\epsilon$ & $\epsK/\epsT$ & $\epsM/\epsT$ \\
\hline
0.02&$0.34\pm0.02$&$0.70\pm0.04$&$0.08\pm0.01$&$0.92\pm0.01$\\
0.10&$0.32\pm0.03$&$0.66\pm0.08$&$0.24\pm0.03$&$0.81\pm0.02$
\label{Tsum2}\end{tabular}}\end{table}

\subsection{Spectral properties and energy dissipation}

\begin{figure}[b!]\begin{center}
\includegraphics[width=\columnwidth]{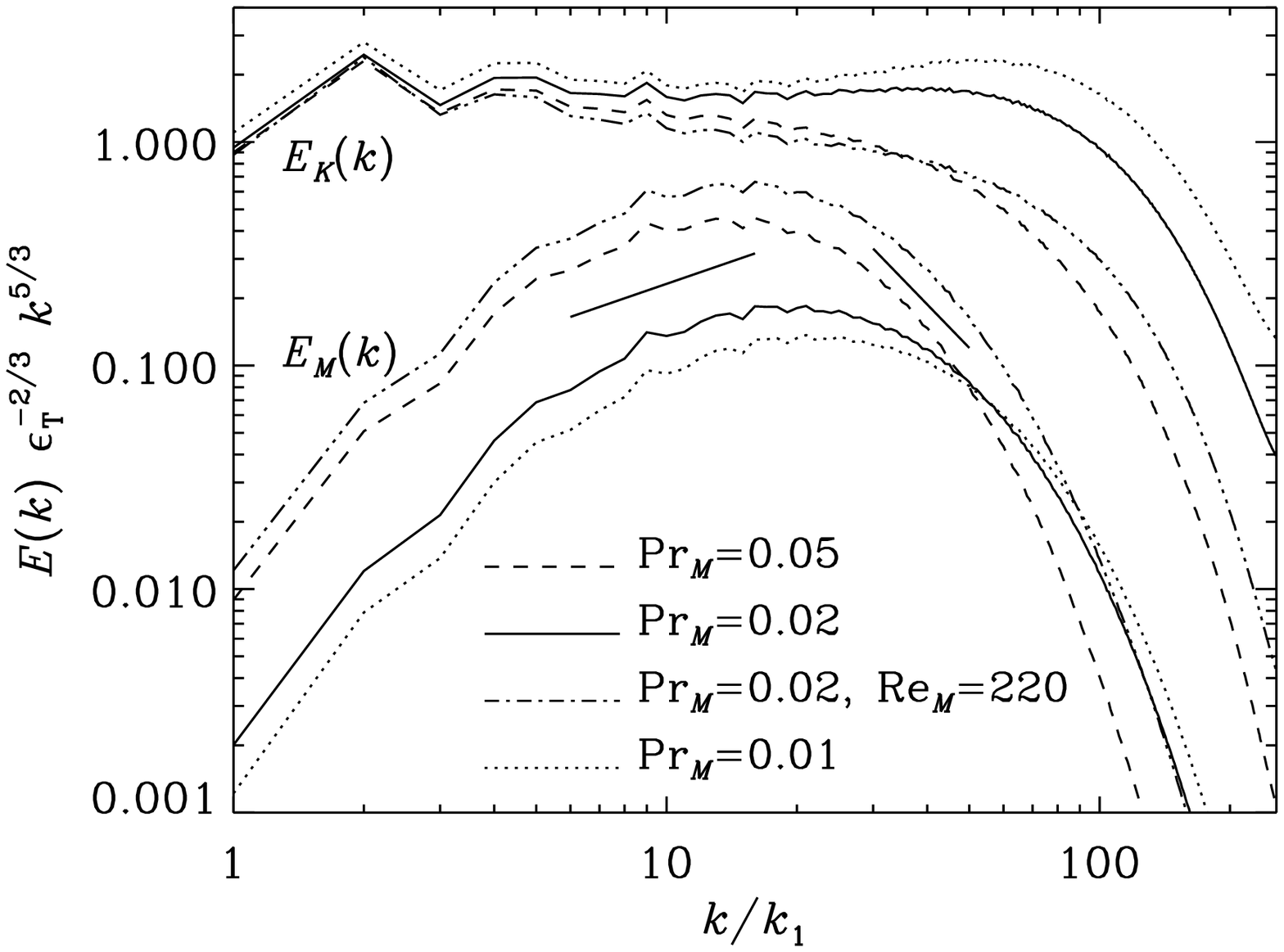}
\end{center}\caption[]{
Compensated kinetic and magnetic energy spectra for runs
with $\Pm=0.05$, $\Pm=0.02$, and $\Pm=0.01$ for $\Rm\approx150$
as well as one run with $\Pm=0.02$ and $\Rm\approx220$.
The resolution is in all cases $512^3$ mesh points.
The two short straight lines give, for comparison, the slopes
$2/3$ (corresponding to a $k^{-1}$ spectrum for $k<20k_1$) and
$-2$ (corresponding to a $k^{-11/3}$ spectrum for $k>20k_1$).
}\label{pspec2}\end{figure}

We consider here kinetic and magnetic energy spectra,
$\EK(k)$ and $\EM(k)$, respectively.
They are normalized in the usual way such that
$\int\EK\,\dd k=\half\rho_0\bra{\UU^2}$ and 
$\int\EM\,\dd k=\half\mu_0^{-1}\bra{\BB^2}$.
In \Fig{pspec2}, these spectra are compensated with
$\epsT^{-2/3}k^{5/3}$.
For $\Pm=0.02$ and 0.01, the kinetic energy spectra show a clear
bottleneck effect, i.e., there is a weak uprise of the compensated spectra
toward the dissipative subrange \citep{Fal94,Kan03,Dob03}.
The compensated magnetic energy spectra peak around $k=20k_1$.
Both toward larger and smaller values of $k$ there is no clear
power law behavior, although the slopes of the $k^{-11/3}$ spectrum
of \cite{Gol60,Mof61} and the scale-invariant $k^{-1}$ spectrum
\citep{RS82,KR94,KMR96} are shown for comparison.

\begin{figure}[h!]\begin{center}
\includegraphics[width=\columnwidth]{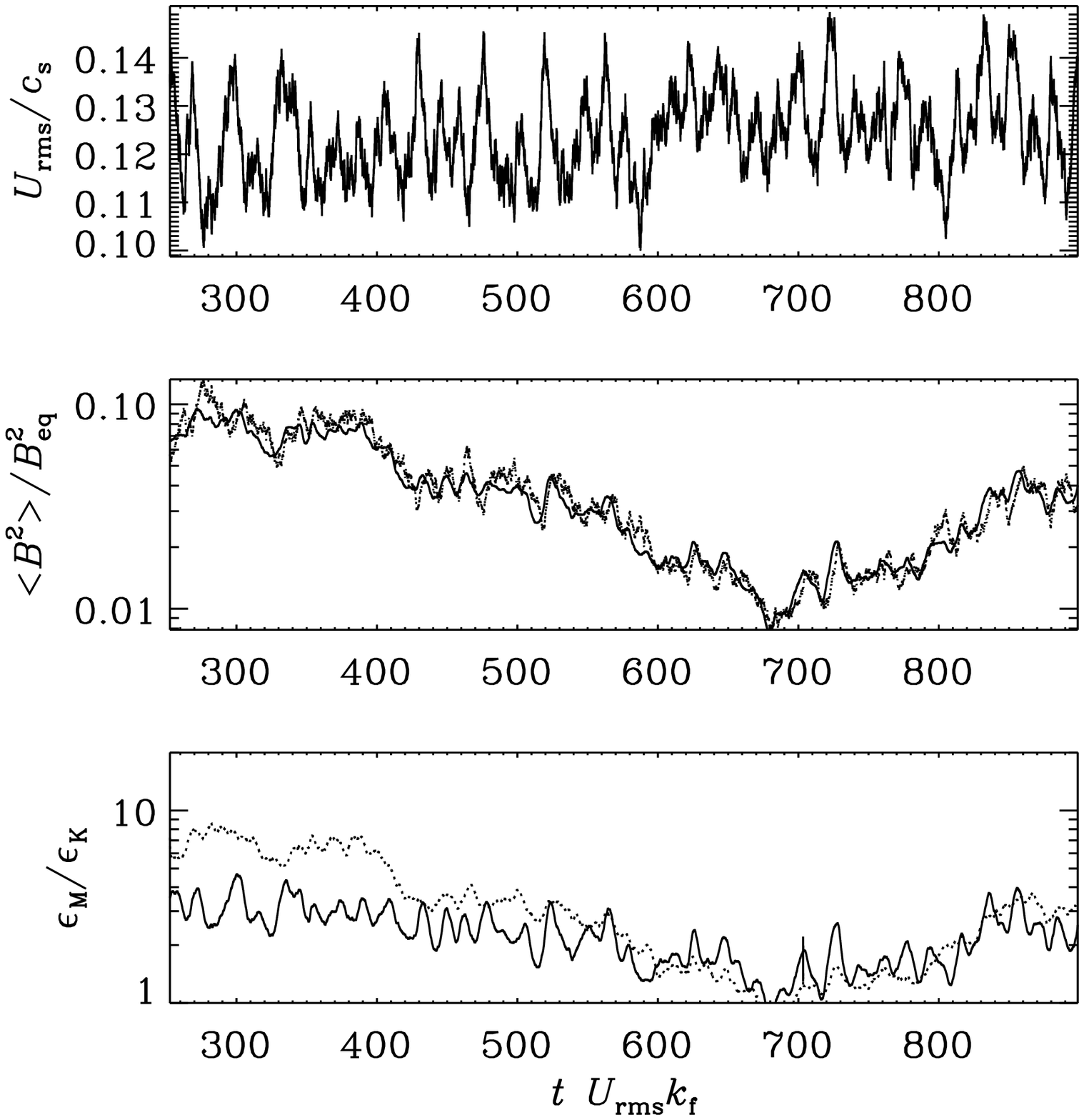}
\end{center}\caption[]{
Root-mean-square velocity, ratio of magnetic to kinetic energy,
as well as the ratio of magnetic to kinetic energy dissipation
for the run with $\Pm=0.02$ using $512^3$ mesh points.
In the last two panels the dashed lines denote normalization
with respect to the instantaneous values of $\Beq$ and $\epsK$,
respectively, while the solid lines refer to normalizations
based on the time averaged values of $\Beq$ and $\epsK$.
}\label{pdiss}\end{figure}

It turns out that for small values of $\Pm$, dynamo action is maintained
for $\Rm\approx160$, corresponding to $\Rey\approx7800$.
This value of $\Rey$ is rather large for a resolution of $512^3$
mesh points and one must be concerned about insufficient resolution.
Similar circumstances were encountered previously in connection with
simulations of large-scale dynamos at low values of $\Pm$ \citep{B09},
and even at large values of $\Pm$ \citep{B11}.
In the former case, much of the energy dissipation occurs magnetically
via Joule dissipation, leaving thus very little energy in the rest of
the kinetic energy cascade.
This allows us then to decrease $\nu$ further, while still allowing
the remaining kinetic energy to get dissipated.
However, kinetic and magnetic energies are quite intermittent (uppermost
panel of \Fig{pdiss}) and there can be extended periods over which the
magnetic energy drops well below the kinetic energy.
Nevertheless, the magnetic energy dissipation is still in excess of the
kinetic energy dissipation; see \Fig{pdiss}.

For $\Pm=0.01$, the nominal value of $\Rey$ is 16,000.
The kinetic energy spectrum extends now further to higher wavenumbers,
but it shows still a monotonic decrease down to the Nyquist wavenumber
at $k=256k_1$.
As can be seen from \Tab{Tsum}, the nominal dissipation wavenumber, $\kK$,
is now well outside the range of resolved wavenumbers, so it is clear that
higher resolution would be needed to resolve the smallest scales properly.
However, as far as the dynamo is concerned, most of the magnetic
field generation occurs at wavenumbers below $30k_1$, which is
where the compensated magnetic energy spectrum begins to show a
clear decline into the dissipation subrange.
Until that wavenumber, significant amounts of kinetic energy are
being channeled into magnetic energy via the dynamo, which lowers
the kinetic energy dissipation and is the main reason for being
able to run such low $\Pm$ cases.

\begin{figure}[t!]\begin{center}
\includegraphics[width=\columnwidth]{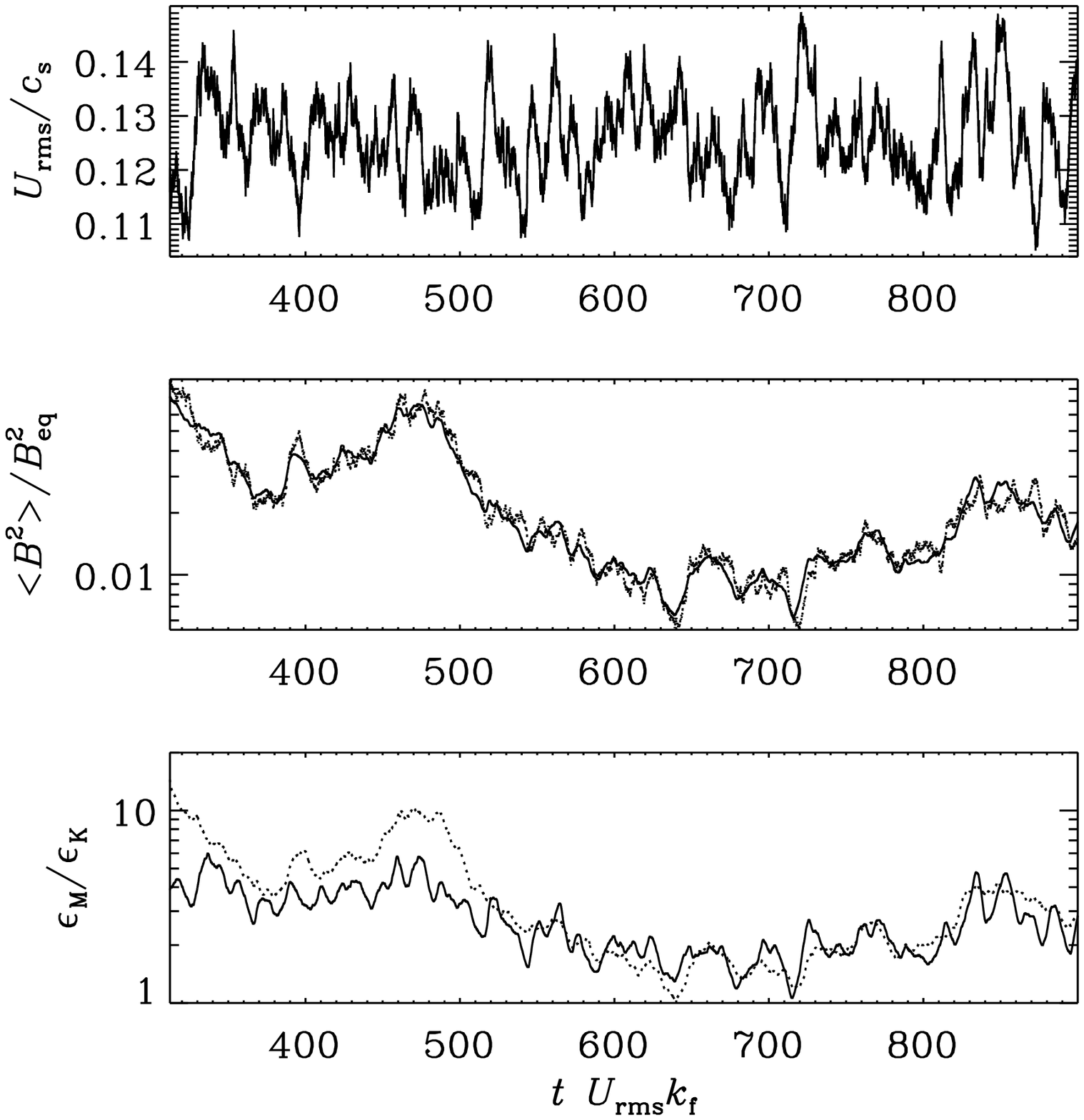}
\end{center}\caption[]{
Same as \Fig{pdiss}, but for the run with $\Pm=0.01$.
}\label{pdiss_512_Pm001a}\end{figure}

The velocity field is relatively steady over the course of the
simulation, but the magnetic field and also the magnetic energy
dissipation vary significantly; see \Fig{pdiss_512_Pm001a}.
However, although there can occasionally be a dramatic decline
in the magnetic field, it tends to recover subsequently, suggesting
that dynamo action is still possible at small values of $\Pm$.
Obviously, in addition to longer run times, it is necessary to perform
simulations at higher resolution, which is not currently feasible
if one wants to cover sufficiently many turnover times.

\begin{figure}[t!]\begin{center}
\includegraphics[width=\columnwidth]{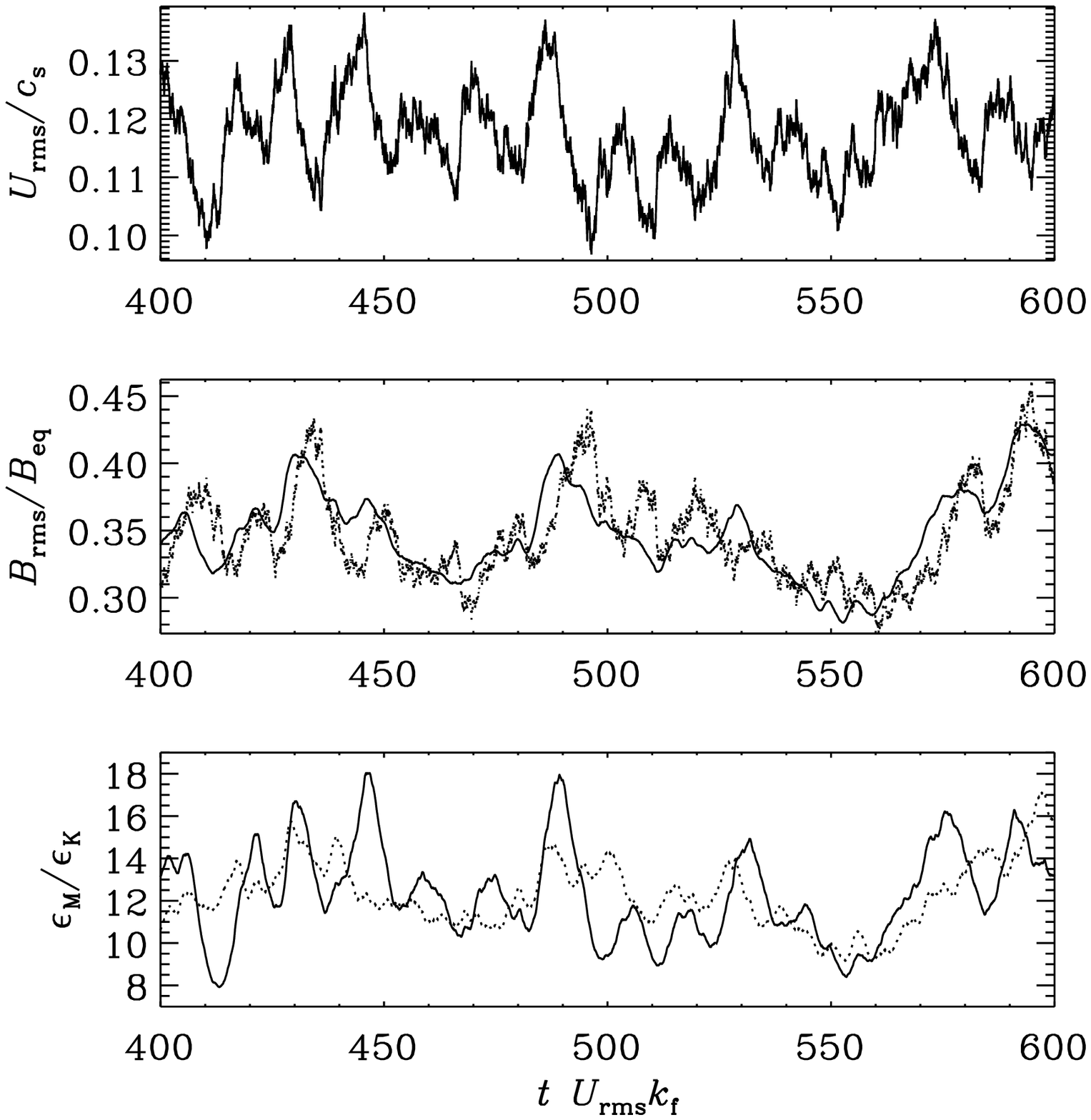}
\end{center}\caption[]{
Similar to \Fig{pdiss}, but for the run with $\Rm=220$,
using still $\Pm=0.02$.
}\label{pdiss_512_Pm002b}\end{figure}

We have already argued that the somewhat erratic behavior of the dynamo at
$\Rm=160$ is a consequence of being close to the marginal value.
The time evolution for the case with $\Rm=220$ and $\Pm=0.02$ is
shown in \Fig{pdiss_512_Pm002b}.
Both $\Brms$ and $\epsK$ are now much closer to being statistically steady.
Furthermore, the value of $\Brms/\Beq$ is now $\approx0.3$ both for
$\Pm=0.1$ and for $\Pm=0.02$.
This suggests that the saturation level of the dynamo is now
beginning to be independent of the value of $\Pm$.

\begin{figure}[t!]\begin{center}
\includegraphics[width=\columnwidth]{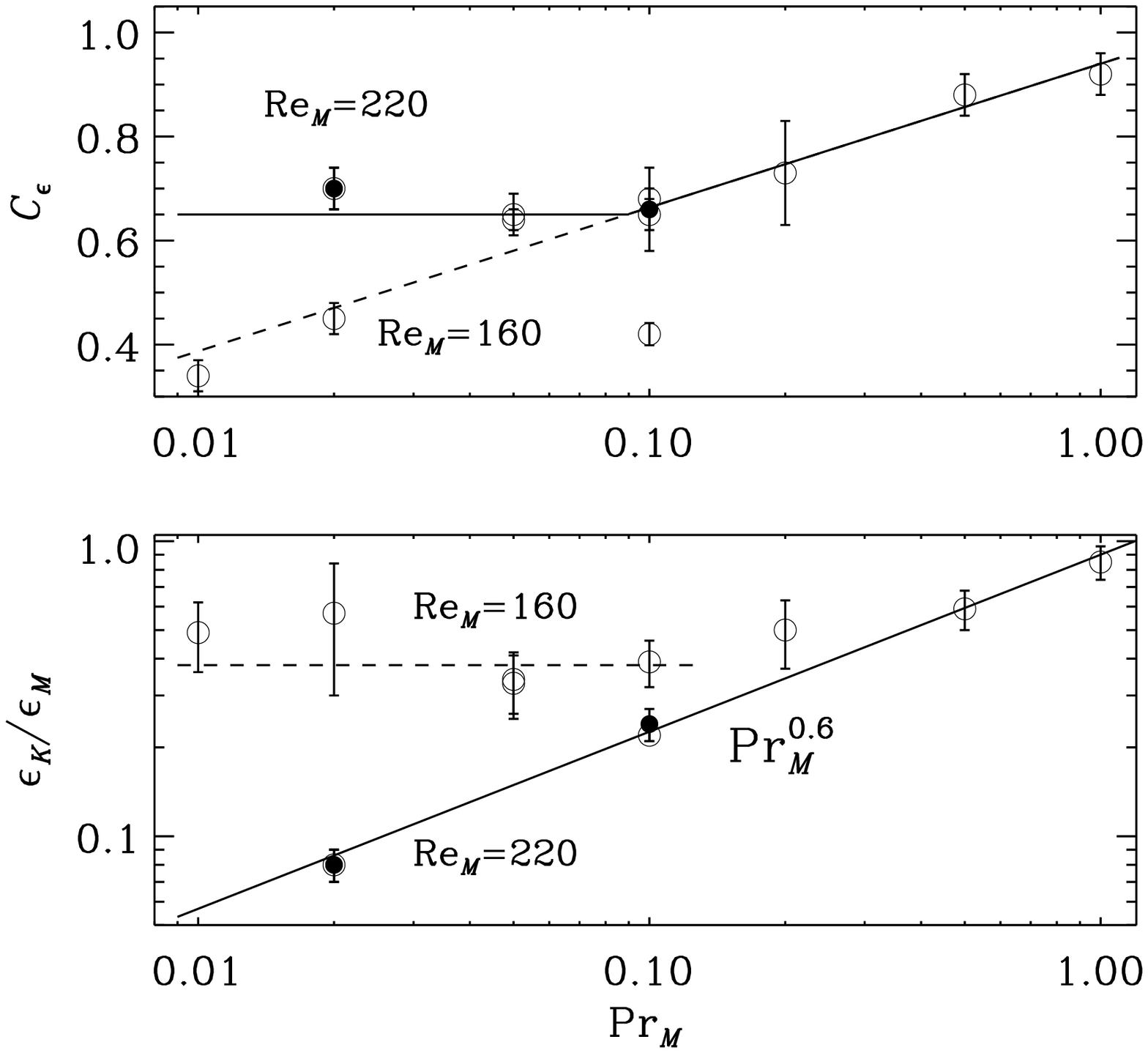}
\end{center}\caption[]{
$\Pm$ dependence of the dimensionless dissipation rate, $C_\epsilon$,
and the kinetic to magnetic energy dissipation ratio, $\epsK/\epsM$.
}\label{peps_vs_Pm}\end{figure}

Earlier work on large-scale dynamos from helical isotropic
turbulence showed that the total magnetic energy dissipation
is larger than in hydrodynamic turbulence.
This is best demonstrated by considering the
conventionally defined dimensionless dissipation parameter
\EQ
C_\epsilon={\epsilon_T\over U^3/L},
\EN
where $U$ is the one-dimensional rms velocity, which is related to
$\urms$ via $U^2=\urms^2/3$, and $L$ is the integral scale which is
related to $\kf$ via ${3\over4}\pi/\kf$.
In non-helical turbulence this value is typically around 0.5
\citep[see, e.g.][]{Pearson}, but in helical turbulence with large-scale
dynamo action this value is around 1.4; see also \cite{B11}.
In \Fig{peps_vs_Pm} we use time averaged dissipation rates, which,
for simplicity, are also denoted by $\epsK$, $\epsM$, and $\epsT$.
In the upper panel of \Fig{peps_vs_Pm} we show that, in the
present case of small-scale dynamo action from non-helical
isotropic turbulence, this value is now closer to the hydrodynamic
value and is slightly above 0.6 for $\Pm=0.02$ and $\Rm=220$;
see upper panel of \Fig{peps_vs_Pm}.
In the lower panel of \Fig{peps_vs_Pm} we see that the ratio
$\epsK/\epsM$ is compatible with a $\Pm^{0.6}$ dependence, as was
found earlier for helical hydromagnetic turbulence \citep{B09,B11}.
However, for $\Rm=160$, $\epsK/\epsM$ levels off at a constant value
of $\approx0.4$, which is probably an artifact of $\Rm$ being too
close to the onset of dynamo action.

\subsection{Possibility of subcritical dynamo action}

We recall that we have used here the strategy of generating
low-$\Pm$ solutions by gradually decreasing $\nu$, and hence
increasing the value of $\Rey$.
As in the case of helical dynamos \citep{B09}, the fact that
a turbulent self-consistently generated magnetic field is present
helps reaching these low-$\Pm$ solutions.
However, the presence of the magnetic field also modifies the
kinetic energy spectrum and makes it decline slightly more steeply
than in the absence of a magnetic field; see \Fig{pspec2}.
This suggests that the velocity field would be less rough
than in the corresponding case without magnetic fields.
Following the reasoning of \cite{BC04}, this should make
the dynamo more easily excited than in the kinematic case
with an infinitesimally weak magnetic field.
In other words, there is the possibility of a subcritical bifurcation
where the dynamo requires a significantly larger value of $\Pm$ to
bifurcate from the trivial $\BB=\bm{0}$ solution than the value
needed to sustain a saturated dynamo.

\begin{figure}[t!]\begin{center}
\includegraphics[width=\columnwidth]{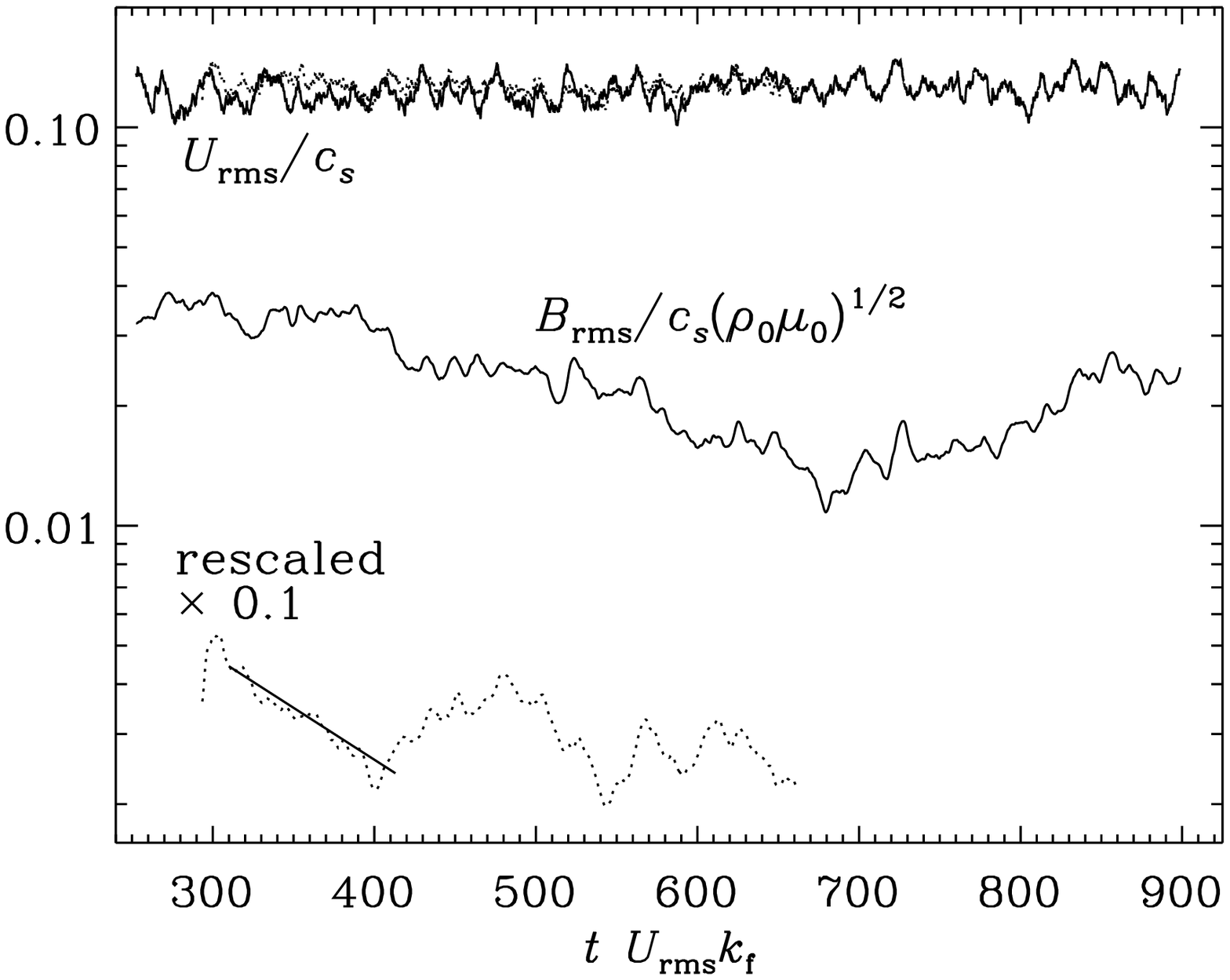}
\end{center}\caption[]{
Evolution of magnetic and kinetic energies in the main run (solid lines)
and after rescaling the magnetic field by a factor of one tenth (dotted lines).
In that case, the resulting decay rate is $0.006\urms\kf$.
}\label{pcomp}\end{figure}

In order to check this hypothesis, we perform an experi\-ment where the
simulation is continued after having down-scaled the magnetic field by
a factor of 10.
The result is shown in \Fig{pcomp}, where we compare the original
simulation with the one restarted with a 10 times lower field.
One sees a gradual decline of the magnetic field after
a brief initial increase of the magnetic field.
This initial increase is a consequence of the reduced feedback from the
Lorentz force, allowing the velocity to increase slightly above
the previous value (see the upper dotted line in \Fig{pcomp}).
During the next 100 turnover times,
the decay rate is about $0.006\urms\kf$, which is about 4 times
smaller than the growth rate of $0.025\urms\kf$ for a non-helical dynamo
at $\Pm=1$ and $\Rey\approx150$ \citep{HBD04}.
However, the field still seems to recover and shows in the end a behavior
comparable to that without rescaling.
This may suggest that at this value of $\Pm$ the dynamo may not be
subcritical after all.

The possibility of subcritical dynamo action is well known in the
geodynamo context, where the flow is driven by thermal or compositional
convection \citep{Rob88}, and for Keplerian shear flows \citep{Rincon}.
Also in the context of dynamos from forced Taylor-Green flows the
possibility of subcritical dynamos is well known \citep{Ponty}.
In the present context, subcriticality is likely to be linked
to the steeper kinetic energy spectrum in the low-$\Pm$ regime.
However, because of extended transients, the results for $\Pm=0.02$
shown in \Fig{pcomp} remain inconclusive.
For $\Pm=0.1$, on the other hand, \cite{Scheko07} have not seen
dynamo action in the linear regime when $\Rm=160$.

\section{Conclusions}

In the present paper we have extended the work of \cite{Iska07} and
\cite{Scheko07} to the nonlinear regime of a saturated dynamo.
However, while in the former (linear) case the dynamo shows signs of a
depression in the range $0.1\leq\Pm\leq0.2$, the nonlinear saturated dynamo
is found to operate nearly unimpededly in the range $0.02\leq\Pm\leq1$.
Furthermore, unlike the work of \cite{Iska07} and \cite{Scheko07}, who used
hyperviscosity, we have here used regular viscosity with the usual
Laplacian diffusion operator.
As in earlier work on helical large-scale dynamos \citep{B09}, it is
possible to reach the regime of low $\Pm$ by restarting the simulations
from another one at a larger value of $\Pm$, which reduces the kinetic
dissipation rate proportional to the square root of $\Pm$.
Furthermore, in contrast to helical large-scale dynamos, where dynamo
onset is possible for values of $\Rm$ of the order of unity and independently
of the value of $\Pm$, we have here the situation where the critical value
of $\Rm$ may be larger than the value required to sustain the dynamo once
it has saturated.
This means that the dynamo could be subcritical and might possesses a
finite amplitude instability at $\Rm$ below and around 160.

We note that the $\Pm^{1/2}$ scaling of the kinetic to magnetic energy
dissipation ratio, $\epsK/\epsM$, is still not well understood.
In view of the definitions of $\epsK$ and $\epsM$ in \Eq{epsdef},
it is clear that this implies that
\EQ
{\nu^{1/2}\bra{\rho\WW^2}\over\eta^{1/2}\bra{\mu_0\JJ^2}}=\const.
\EN
This means that the usually expected scaling for hydrodynamic
turbulence, $\nu\bra{\rho\WW^2}=\const$, or the hydromagnetic scaling
$\eta\bra{\mu_0\JJ^2}=\const$, which has been confirmed for $\Pm=1$
\citep[see Fig.~8 of][]{Candel}, is clearly not generally valid and
needs to be reconsidered.

Our results for the magnetic energy spectra are consistent with those of
earlier direct numerical simulations by \cite{Scheko07} in that there
is a short Golitsyn $k^{-11/3}$ spectrum near the resistive scale,
as well as a short $k^{-1}$ spectrum on larger scales.
Both properties have also been seen in liquid sodium experiments
\citep{OPF98,Bou02} as well as in large eddy simulations \citep{Ponty04}.

The astrophysical relevance of small-scale dynamo action is hardly
disputed.
Even in situations were large-scale dynamo action is possible, like in
the Sun, small-scale magnetic fields are seen ubiquitously even in the
quiet photosphere where there is no evidence of any effects from the
large-scale field \citep{Cat99,VS07,Piet10,Jin11}.
The present work now confirms that the small value of the Sun's magnetic
Prandtl number may not be a problem with this proposal.
Although one may worry that most of the simulations presented
so far have overestimated the effects of small-scale dynamo action by
having chosen values of $\Pm$ of the order of unity \citep{B05},
it is remarkable that the field strength decreases only slightly
when we decrease $\Pm$ from 1 to 0.02, provided $\Rm$ is large enough
($\Rm\ga150$).
Future work will hopefully clarify further the relative importance of
large-scale and small-scale dynamo action in astrophysical bodies like
the Sun.

Another aspect that needs to be addressed in future simulations concerns
the magnetic Prandtl number effects on the large-scale properties of
the turbulence.
This concerns in particular the turbulent diffusion of large-scale
magnetic fields, as can be measured by the quasi-kinematic test-field
method, for example \citep{BRRS08}.
Among other things, one would like to confirm that the turbulent magnetic
diffusivity is not affected by the small-scale magnetic field,
which is a standard result in mean-field theory \citep{GD94,RKR03,BS05}.
In that case, if $\Rm$ is close to the onset of small-scale dynamo action,
one would expect the turbulent diffusion to be independent of the value
of $\Pm$.
However, for large values of $\Rm$, as we have now seen, the effect of $\Pm$
on the small-scale dynamo is less dramatic.
Thus, even if the turbulent magnetic diffusivity was affected by the
small-scale field, the effect could only be weak.

\acknowledgements
I thank Paul Roberts for hospitality during a visit to Malibu where
this project was started.
I also acknowledge the organizers of the KITP program on turbulence
for providing a stimulating atmosphere.
This research was supported in part by the National Science Foundation
under grant PHY05-51164 and the European Research Council under the
AstroDyn Research Project 227952.
The computations have been carried out at the National Supercomputer
Centre in Ume{\aa} and at the Center for Parallel Computers at the
Royal Institute of Technology in Sweden.


\end{document}